\documentstyle[preprint,aps]{revtex}  
\begin{document}                
\preprint{KNUTH-26, \\March \\1995}
\draft

\title{Unitarity Condition on
Quantum Fields in Semiclassical Gravity}
\author{Sang Pyo Kim \footnote{
E-mail : sangkim@knusun1.kunsan.ac.kr
}}
\address{Department of Physics \\ Kunsan National University \\
Kunsan 573-701, Korea}

\maketitle
\begin{abstract}                
The condition for the unitarity of a quantum field is
investigated in semiclassical gravity from
the Wheeler-DeWitt equation.
It is found that
the quantum field  preserves unitarity asymptotically
in the Lorentzian universe, but does not
preserve unitarity completely
in the Euclidean universe.
In particular we obtain a very simple
matter field equation in the basis of the generalized invariant
of the matter field Hamiltonian whose asymptotic
solution is found explicitly.
\end{abstract}
\medskip
\medskip
\medskip
\medskip
\medskip
\medskip
{\centerline {\it Published in Physics Letters A {\bf 205}, 359 (1995)}}
\medskip
\medskip
\medskip
\medskip
\medskip
\medskip

Unitarity of quantum field theory in curved space-time has
been a problem long debated but sill unsolved.
In particular the issue has become an impassioned
altercation with the discovery of
the Hawking radiation \cite{Hawking} from black hole
in relation to the information loss problem.
Recently there has been a series of
active and intensive investigations
of quantum effects of matter field through dilaton gravity
and resumption of unitarity and information loss
problem (for a good review and references see \cite{Page}).

In this letter we approach the unitarity
problem and investigate the condition for the unitarity
of a quantum field
from the point of view of semiclassical
gravity based on the Wheeler-DeWitt equation \cite{Banks}.
By developing various methods
\cite{Brout1,Kiefer1,Singh1,Brout2,Singh2,Balbinot,Kiefer2,Paz,Gundlach,Kiefer3,Datta,Kim1,Kim2,Kiefer4,Kim3}
for semiclassical gravity and elaborating further the new
asymptotic expansion method \cite{Kim4} for the
Wheeler-DeWitt equation,
we derive the quantum field theory for a matter
field from the Wheeler-DeWittt equation
for the gravity coupled to the matter field, which
is equivalent to a gravitational field equation and
a matrix equation for the matter field through
a definition of cosmological time. The full
field equation for the matter field is found to
preserve unitarity asymptotically
in the limit $\frac{\hbar}{M} \rightarrow 0$
in the Lorentzian universe with an oscillatory
gravitational wave function
but does not preserve unitarity completely for an exponential
gravitational wave function in
the Euclidean universe.
We find the exact quantum state for the asymptotic field equation
in terms of the eigenstates of the generalized invariant
of the matter field Hamiltonian.

We consider the Wheeler-DeWitt equation
for a quantum cosmological model
\begin{equation}
\left[ - \frac{\hbar^2}{2M} \nabla^2 - MV(h_a) +
\hat{H}(\frac{i}{\hbar} \frac{\delta}{\delta \phi}, \phi, h_a)
\right] \Psi(h_a,\phi) = 0.
\label{WD eq.}
\end{equation}
Here $M$ is the Planck mass squared,
$\nabla^2 = G_{ab} (\delta^2)/(\delta h_a \delta h_b)$,
$V$ denotes the superpotential of three-curvature with or without
the cosmological constant
on superspace with the DeWitt metric $G_{ab}$
with the signature $(-,+,\cdots,+)$,
and $\hat{H}$ represents the matter field Hamiltonian.
We find the wave function of the form
\begin{equation}
\Psi (h_a,\phi) = \psi (h_a) \Phi (\phi,h_a),
\label{wave func.}
\end{equation}
where $\Phi (\phi,h_a)$ is a gravitational
field-dependent quantum state
of the matter field. We expand the quantum
state by some orthonormal basis
\begin{equation}
\Phi (\phi,h_a) = \sum_{k}^{\atop} c_k (h_a)
\left| \Phi_k(\phi,h_a) \right> , ~~
\left<\Phi_k | \Phi_n \right> = \delta_{kn}.
\label{quantum state}
\end{equation}
Substituting Eqs. (\ref{wave func.})
and (\ref{quantum state}) into
Eq. (\ref{WD eq.}) and acting
$\left< \Phi_n \right|$ on both sides,
one obtains the following matrix equation
equivalent to the Wheeler-DeWitt equation
\begin{eqnarray}
c_n(h_a) \left( - \frac{\hbar^2}{2M} \nabla^2
- MV(h_a) + H_{nn} (h_a) \right) \psi(h_a)
- \frac{\hbar^2}{M} \nabla \psi (h_a) \cdot \nabla c_n (h_a)
\nonumber\\
+ i \frac{\hbar^2}{M} \nabla \psi (h_a) \cdot
\sum_{k}^{\atop}
{\bf A}_{nk} (h_a) c_k (h_a)
+ \psi(h_a) \sum_{k \neq n}^{\atop}
H_{nk} (h_a) c_k(h_a)
\nonumber\\
- \frac{\hbar^2}{2M} \psi (h_a)
\sum_{k}^{\atop} \Omega_{nk} (h_a) c_k (h_a) = 0,
\label{matrix eq.}
\end{eqnarray}
where
\begin{eqnarray}
H_{nk} (h_a) &=& \left< \Phi_n (h_a) \right| \hat{H}
\left| \Phi_k (h_a) \right>,
\nonumber\\
{\bf A}_{nk} (h_a) &=& i \left< \Phi_n (h_a) \right| \nabla
\left| \Phi_k (h_a) \right>,
\nonumber\\
\Omega_{nk} (h_a) &=&
\nabla^2 \delta_{nk} -2i {\bf A}_{nk} \cdot \nabla
+ \Omega_{nk}^{(2)},
\end{eqnarray}
where
\begin{equation}
\Omega_{nk}^{(2)} (h_a) = \left< \Phi_n (h_a) \right| \nabla^2
\left| \Phi_k (h_a) \right>.
\end{equation}
By noting that $i \nabla$ acts on $c_k(h_a)$
as a hermitian operator, it follows
that $H$, ${\bf A}$, $\Omega^{(2)}$,
and thereby $\Omega$ are hermitian matrices.

We separate the matrix equation
equivalent to the Wheeler-DeWitt equation
into the gravitational part
\begin{equation}
\left( - \frac{\hbar^2}{2M} \nabla^2
- MV(h_a) + H_{nn} (h_a) \right) \psi(h_a) = 0,
\label{grav. eq.}
\end{equation}
and the matter field part
\begin{eqnarray}
- \frac{\hbar^2}{M} \nabla \psi (h_a) \cdot \nabla c_n (h_a)
+ i \frac{\hbar^2}{M} \nabla \psi (h_a) \cdot
\sum_{k}^{\atop}
{\bf A}_{nk} (h_a) c_k (h_a)
\nonumber\\
+ \psi(h_a) \sum_{k \neq n}^{\atop}
H_{nk} (h_a) c_k(h_a)
- \frac{\hbar^2}{2M} \psi (h_a)
\sum_{k}^{\atop} \Omega_{nk} (h_a) c_k (h_a) = 0.
\label{mat. eq.}
\end{eqnarray}
The gravitational wave function has an effective
potential $MV - H_{nn}$. The gravitational wave function can
have either a real action in a region of the Lorentzian
universe or an imaginary action
in a region of the Euclidean universe.

First, we consider a region of the Lorentzian universe,
in which the gravitational wave function takes the form
\begin{equation}
\psi(h_a) = f(h_a) \exp\left(\frac{i}{\hbar}
S_{nn} (h_a) \right).
\label{grav. ac.}
\end{equation}
In the semiclassical limit $\hbar \rightarrow 0$,
the gravitational action satisfies the Einstein-Hamilton-Jacobi
equation with the quantum back-reaction
\begin{equation}
\frac{1}{2M} \left(\nabla S_{nn}(h_a) \right)^2
- M V(h_a) + H_{nn} (h_a) = 0.
\end{equation}
There are possibly an infinite number of the
gravitational wave functions depending on the mode number
due to the quantum back-reaction of the matter field.
Each gravitational wave function, whose peak
corresponds to a classical solution with the matter field,
describes a history
of evolution of the universe and contains
the whole information of the universe.
So along each wave function in this region
we may introduce a cosmological time
\begin{equation}
\frac{\partial}{\partial \tau^{(n)}} : =
\frac{1}{M}
\nabla S_{nn} (h_a) \cdot \nabla.
\label{cosmological time}
\end{equation}
Substituting (\ref{grav. ac.}) into (\ref{mat. eq.})
and dividing $\psi(h_a)$, we obtain the
matter field equation
\begin{equation}
i\hbar \frac{\partial}{\partial \tau^{(n)}} c_n
+ \Omega_{nn}^{(1)} c_n
+ \sum_{k \neq n}^{\atop} \left(
\Omega_{nk}^{(1)} - H_{nk} \right) c_k
+ \frac{\hbar^2}{2M}
\sum_{k}^{\atop} \Omega^{(3)}_{nk} c_k = 0,
\label{matter field eq.}
\end{equation}
where
\begin{equation}
\Omega_{nk}^{(1)} = i\hbar \left< \Phi_n \right|
\frac{\partial}{\partial \tau^{(n)}}
\left| \Phi_k \right>
= \frac{\hbar}{M} \nabla S_{nn}
\cdot {\bf A}_{nk},
\end{equation}
and
\begin{equation}
\Omega^{(3)}_{nk} = \Omega_{nk}
+ 2 \frac{1}{f} \nabla f \cdot \nabla
\delta_{nk} - 2i \frac{1}{f} \nabla f \cdot
{\bf A}_{nk}.
\end{equation}
The full field equation (\ref{matter field eq.})
for the matter field is not unitary
due to the terms
$2 \frac{1}{f} \nabla f \cdot \nabla
\delta_{nk} - 2i \frac{1}{f} \nabla f \cdot
{\bf A}_{nk}$ which do not obviously act as unitary
operators,
even though all the other terms $\Omega^{(1)}$,
$H_{nk}$, and $\Omega_{nk}$ are hermitian matrices.
The unitarity violating terms were discovered using
different method in Ref.\cite{Kiefer3}.
In a previous paper \cite{Kim4}, we also obtained
a small unitarity
violating term, $ - i \frac{\hbar}{2M}
\nabla^2 S_{nn}(h_a)$,
which originated from the gravitational
wave function in the form $\psi(h_a)
= \exp(\frac{i}{\hbar} S_{nn}(h_a))$.

However, it is only in
the asymptotic limit $\frac{\hbar}{M} \rightarrow 0$
that for any gravitational wave function satisfying
(\ref{grav. eq.}) these unitarity violating terms
are suppressed and one gets
\begin{equation}
i\hbar \frac{\partial}{\partial \tau^{(n)}} c_n
+ \Omega_{nn}^{(1)} c_n
+ \sum_{k \neq n}^{\atop} \left(
\Omega_{nk}^{(1)} - H_{nk} \right) c_k
= 0.
\label{unit field eq.}
\end{equation}
In this case,
since $H^{\dagger} = H$, $\Omega^{(1)\dagger}
= \Omega^{(1)}$, and
$\Omega^{\dagger} = \Omega$, it follows that
\begin{equation}
\frac{\partial}{\partial \tau^{(n)}}
\left( {\bf c}^{\dagger} \cdot {\bf c}
\right)
= 0,
\label{norm}
\end{equation}
where ${\bf c}$ denotes a column vector of ${c_k}$.
The norm of vector, $|{\bf c}|$, is preserved.
This implies a unitary operator such that
\begin{equation}
{\bf c}(\tau^{(n)}) = {\bf U}_c(\tau^{(n)},\tau^{(n)}_0)
{\bf c}(\tau^{(n)}_0).
\end{equation}
And since the eigenstates of the generalized invariant form
an orthonormal basis, there is also a unitary operator
\begin{equation}
\left| \Phi (\tau^{(n)}) \right> = {\bf U}_{\Phi}
(\tau^{(n)},\tau^{(n)}_0)
\left| \Phi(\tau^{(n)}_0) \right>
\end{equation}
for the evolution of the column vector of
$\left| \Phi_k \right>$.
The physical implication is that
the quantum field theory of the matter field is
asymptotically unitary in this sense.

In particular, in terms of the eigenstates of
the generalized invariant
\begin{eqnarray}
\frac{\partial}{\partial \tau^{(n)}} \hat{I}
- \frac{i}{\hbar} \left[ \hat{I} , \hat{H} \right]
= 0,
\nonumber\\
\hat{I} \left| \Phi_k \right> = \lambda_k
\left| \Phi_k \right>
\end{eqnarray}
there is a well-known decoupling theorem \cite{Lewis}
such that
\begin{equation}
H_{nk} = \Omega_{nk}^{(1)}, ~~ n \neq k.
\end{equation}
In the basis of the eigenstates of the generalized invariant
the matter field equation takes the simpler form
\begin{equation}
i\hbar \frac{\partial}{\partial \tau^{(n)}} c_n
+ \Omega_{nn}^{(1)} c_n
+ \frac{\hbar^2}{2M}
\sum_{k}^{\atop} \Omega^{(3)}_{nk} c_k = 0.
\label{non mf eq.}
\end{equation}
In the asymptotic limit of $\frac{\hbar}{M} \rightarrow \ 0$
we obtain the  asymptotic quantum
state of matter field \cite{Kim3}
\begin{equation}
\Phi(\phi,h_a) =  c_n (\tau^{(n)}_0) \exp
\left( \frac{i}{\hbar}
\int \Omega_{nn}^{(1)} (h_a) d\tau^{(n)} \right)
\left| \Phi_n (\phi,h_a) \right>.
\label{as so}
\end{equation}
With an additional phase factor $\exp \left(-\frac{i}{\hbar}
\int H_{nn} d \tau^{(n)}\right)$
(\ref{as so}) is also one of the exact quantum states of
time-dependent Schr\"{o}dinger equation
\begin{equation}
i \hbar \frac{\partial}{\partial \tau^{(n)}}
\Phi(\phi, h_a)
= \hat{H}(\frac{i}{\hbar}
\frac{\delta}{\delta \phi}, \phi,h_a) \Phi(\phi, h_a).
\end{equation}
The exact quantum state (\ref{quantum state})
can be obtained solving pertubativley (\ref{non mf eq.}),
which is a linear superposition of eigenstates of the
generalized invariant with gravitational field-dependent
coefficient functions.

Second, we consider an exponential gravitational wave function
of the form in the Euclidean universe
\begin{equation}
\psi(h_a) = f(h_a) \exp(- \frac{1}{\hbar} S_{nn} (h_a)).
\end{equation}
The cosmological time is imaginary and given by
\begin{eqnarray}
\frac{\partial}{\partial \tau} &=& i \frac{1}{M}
\nabla S_{nn} \cdot \nabla
\nonumber\\
: &=& -i \frac{\partial}{\partial \tau^{(n)}_{im}}.
\end{eqnarray}
The matter field equation becomes
\begin{equation}
\hbar \frac{\partial}{\partial \tau^{(n)}_{im}} c_n
+ \Omega_{nn}^{(1)} c_n
+ \sum_{k \neq n}^{\atop} \left(
\Omega_{nk}^{(1)} - H_{nk} \right) c_k
+ \frac{\hbar^2}{2M}
\sum_{k}^{\atop} \Omega^{(3)}_{nk} c_k = 0.
\label{nonun. matter field eq.}
\end{equation}
Even though the basis evolves unitarily, the coefficient functions
in (\ref{quantum state}) with respect
to which the exact quantum state
is expanded do not. This violates completely
the unitarity of exact quantum state.

As an application, we consider a free
massive scalar field coupled
to the gravity. Through the definition of
cosmological time (\ref{cosmological time})
using the gravitational wave function
the Hamiltonian for the matter field depends on the time as
\begin{equation}
\hat{H}(h_a) = \frac{1}{m(\tau)}
\frac{\pi_{\phi}^2}{2}
+ m(\tau) \omega^2(\tau)
\frac{\phi^2}{2}.
\end{equation}
It is found that the one-parameter-dependent
generalized invariant
\begin{equation}
\hat{I} (\tau) = g_-(\tau) \frac{\pi_{\phi}^2}{2}
+ g_0(\tau) \frac{\pi_{\phi} \phi + \phi \pi_{\phi}}{2}
+ g_+(\tau) \frac{\phi^2}{2},
\end{equation}
is given by
\begin{eqnarray}
g_-(\tau) &=& c_1 \left( \phi_1^2 (\tau)
+ \phi_2^2 (\tau) \right),
\nonumber\\
g_0(\tau) &=& c_1 m(\tau) \left( \phi_1 (\tau)
\dot{\phi}_2 (\tau)
+ \phi_2 (\tau) \dot{\phi}_1 (\tau) \right),
\nonumber\\
g_+(\tau) &=& c_1 m^2(\tau) \left( \dot{\phi}_1^2 (\tau)
+ \dot{\phi}_2^2 (\tau) \right)
\end{eqnarray}
in the Lorentzian universe \cite{Kim5,Ji},
and by
\begin{eqnarray}
g_-(\tau) &=& c_1 \left( \phi_1^2 (\tau)
- \phi_2^2 (\tau) \right),
\nonumber\\
g_0(\tau) &=& i c_1 m(\tau) \left( \phi_1 (\tau)
\dot{\phi}_2 (\tau)
- \phi_2 (\tau) \dot{\phi}_1 (\tau) \right),
\nonumber\\
g_+(\tau) &=& - c_1 m^2(\tau) \left( \dot{\phi}_1^2 (\tau)
- \dot{\phi}_2^2 (\tau) \right)
\end{eqnarray}
in the Euclidean universe \cite{Kim6},
where
\begin{equation}
\ddot{\phi}_{1,2} (\tau) + \frac{\dot{m}(\tau)}{m(\tau)}
\dot{\phi}_{1,2} (\tau)
+ \omega^2 (\tau) \phi_{1,2} (\tau) = 0.
\end{equation}
It should be remarked that
the generalized invariant for a harmonic oscillator
in the Euclidean universe leads to an upside-down oscillator.

Finally, we apply the new method to the well-known
one-dimeniosnal Klein-Gordon equation
\begin{equation}
\left[- \frac{\hbar^2}{c^2}
\frac{\partial^2}{\partial t^2}
+ \hbar^2
\frac{\partial^2}{\partial x^2} - m^2c^2
\right] \Psi (t,x) = 0.
\end{equation}
Here the asymptotic parameter is $c$ the speed of light.
Since the Hamiltonian $\hat{H} = \frac{\partial^2}{\partial x^2}$
is independent of parameter time $t$, we may take the
generalized invariant $\hat{I} = \hat{H}$.
The eigenstates of the generalized invariant are
\begin{equation}
\left|\Phi_k(\phi) \right> = \exp(\pm ikx).
\end{equation}
Not only the gauge potential ${\rm A}_{nk}$ but also
$\Omega_{nk}$ do vanish. The expectation value gives rises
to $H_{nk}  = - \hbar^2k^2 \delta_{kn}$.
The equation that corresponds to
Eq. (\ref{grav. eq.}) becomes
\begin{equation}
\left[ \frac{\hbar^2}{c^2} \frac{\partial^2}{\partial t^2}
+ \hbar^2 k^2 + m^2c^2 \right] \psi(t)
= 0,
\end{equation}
whose solution is
\begin{equation}
\psi (t) = \exp(\pm i\omega t), ~~
\omega = \sqrt{k^2c^2 + \frac{m^2c^4}{\hbar^2}}.
\label{rel freq.}
\end{equation}
It is to be noted that the new method introduced
in this paper gives directly  the exact result
at the asymptotic limit. This argument can also
be verified for a gravitational
field-independent $\hat{H}(i\frac{\delta}{\hbar \delta \phi}, \phi)$.

In this letter as a semiclassical gravity
we recovered an asymptotic unitary theory
of the matter field in curved space in the Lorentzian universe
and a non-unitary theory in the Euclidean universe,
even though the quantum
gravity based on the Wheeler-DeWitt
equation is not a complete theory.
It is expected that the unitarity of quantum field of matter field
may be restored for any complete theory of quantum gravity.

Quite recently Kiefer \cite{Kiefer4} has shown that
the quantum field theory derived from the Wheeler-DeWitt
equation is non-unitary in general and Kiefer {\it et al}
\cite{Kiefer3} have suggested
its implication in black hole evaporation.
The big difference between this paper and theirs
is that we used the new method in which one separates
the Wheeler-DeWitt equation into the gravitational
wave equation with the quantum back-reaction of the
matter field
and the matrix equation for the matter field,
whereas they used the conventional method in
which one separates the Wheeler-DeWitt equation
into the Einstein-Hamilton-Jacobi equation
for the gravitational field
without the quantum back-reaction of the matter field
and the time-dependent functional Schr\"{o}dinger
equation for the matter field including higher
order quantum gravitational correction. As shown
in the Klein-Gordon equation in a flat space-time
our method gives the relativistic frequency
(\ref{rel freq.}) directly
at the lowest order whereas the conventional
method would expand the energy as
$E = mc^2 + \frac{p^2}{2mc^2} - \frac{p^4}{8m^3c^2} + \cdots$.
The difference between the new method and the conventional
method was also discussed in various
quantum cosmological models
in Ref. \cite{Kim4}.

The author would like to thank
S. Carlip, S. -W. Kim, D. La, and K. Maeda
for useful discussions.
This work was supported by the Korea Science
and Engineering Foundation under Grant No 951-0207-056-2.

\end{document}